\documentclass[proceedings]{stacs}
\stacsheading{2009}{553--564}{Freiburg}
\firstpageno{553}

\usepackage{amsmath,amssymb,graphicx
}
\theoremstyle{plain}\newtheorem{fact}[thm]{Fact}

\begin{document}

\title[Ambiguity and Communication]{Ambiguity and Communication}

\author[lab1]{J. Hromkovic}{Juraj Hromkovi\v c}
\address[lab1]{Department of Computer Science, ETH Z\"urich,
\newline ETH Zentrum, CH-8022 Z\"urich, Switzerland}
\email{juraj.hromkovic@inf.ethz.ch}

\author[lab2]{G. Schnitger}{Georg Schnitger}
\address[lab2]{Institut f\"ur Informatik, Goethe Universit\"at,
\newline Robert Mayer-Strasse 11-15, D-6054 Frankfurt a. M., Germany}
\email{georg@thi.informatik.uni-frankfurt.de}

\thanks{Supported by SNF-grant 200020-120073 and DFG-grant SCHN 503/4-1.
Part of the work was done while the second author was visiting the ETH
Z\"urich.}



\begin{abstract}  The ambiguity of a nondeterministic finite automaton
(NFA) $N$ for input size $n$ is the maximal number of accepting computations
of $N$ for an input of size $n$.  For all $k,r \in \mathbb{N}$ we construct
languages $L_{r,k}$ which can be recognized by NFA's with size
$k \cdot$poly$(r)$ and ambiguity $O(n^k)$, but $L_{r,k}$ has only NFA's
with exponential size, if ambiguity $o(n^k)$ is required. In particular, a
hierarchy for polynomial ambiguity is obtained, solving a long standing open
problem (Ravikumar and Ibarra, 1989, Leung, 1998).
\end{abstract}

\maketitle





\section{Introduction}

The ambiguity of an NFA $N$ measures the degree of nondeterminism 
employed by $N$ as a function of the input size: let ${\rm{ambig}}_N(x)$ 
be the number of accepting computations of $N$ on input $x$ and define
$${\rm{ambig}}_N(n) = \max \{ {\rm{ambig}}_N (x) : x \in \Sigma^n
\}$$ to be the ambiguity of $N$.  There are related complexity
measures such as the advice and the leaf complexity of $N$.  To
describe their definition let $T_{N}(x)$ be the computation tree of
$N$ on input $x$.  Then advice$_N (x)$ is the maximum, over all
paths in $T_N(x)$ from the root to a leaf, of the number of nodes
with at least two children and
$${\rm{advice}}_N(n) = \max \{ {\rm{advice}}_N (x) : x \in \Sigma^n\}$$
is the advice complexity of $N$. The leaf complexity of $N$
determines the maximal number of computations for inputs of length
$n$.  Thus, if leaf$_N(x)$ is the number of leaves of $T_{N}(x)$,
then $${\rm{leaf}}_N(n) = \max \{ {\rm{leaf}}_N (x) : x \in \Sigma^n
\} .$$ For a minimal NFA $N$ these measures are related as follows
\cite{HKKSS}
$${\rm{advice}}_N (n), {\rm{ambig}}_N(n) \leq {\rm{leaf}}_N(n) =
O({\rm{advice}}_N(n) \cdot {\rm{ambig}}_N(n))$$ and, since
${\rm{advice}}_N (n)$ is at most linear, leaf complexity and
ambiguity are polynomially related, provided both are at least
linear. Since leaf complexity is either bounded by a constant or at
least linear but polynomially bounded, or otherwise exponential in
the input length, we obtain that ambiguity is either bounded by a
constant or bounded by a polynomial or at least exponential \cite
{HKKSS}.

Advice and leaf complexity are rather coarse measures, since advice
and leaf complexity of an unambiguous NFA may be linear.  Ambiguity
on the other hand also influences the tractability of algorithmic
questions.  For instance, for any fixed $k \in \mathbb{N}$ it can be
determined efficiently whether two NFA's of ambiguity at most $k$ are
equivalent, resp. whether the ambiguity
of a given NFA is at most $k$ \cite{SH85}.

How large is the decrease in conciseness, i.e., the increase in the
number of states, if ambiguity is restricted? To study this
question, four classes of NFA's, namely UNA (unambiguous
nondeterministic automata), FNA (finitely ambiguous NFA), PNA
(polynomially ambiguous NFA) and ENA (exponentially ambiguous NFA)
are introduced in \cite{RI}.  The classification into
FNA's, PNA's or ENA's can be performed efficiently \cite{WS}.

Remember that the ambiguity of an NFA $N$ is either at least
exponential or at most polynomial and hence an NFA is either a PNA
or an ENA. Leung \cite{L98} shows that there are ENA's $N_n$ with
$n$ states such that any equivalent PNA has at least $2^n-1$ states.
Hence ENA's can be far more succinct than PNA's.  Subsequently a
similar result, applicable to a larger class of languages, was shown
in \cite{HKKSS} by using methods of communication complexity. In
particular, the conciseness problem for PNA's can be reduced to the
following communication result for the iterated language of
non-disjointness.  Let $\Sigma_r$ be the alphabet of all
subsets of $\{1, \ldots, r^{32}\}$ of size $r$ and set
$$L_r = \{xy | x,y \in \Sigma_r \mbox{
and } x \cap y \not= \emptyset \}.$$  Thus $(L_r)^t $ consists
of all strings $x_1y_1 \cdots x_ty_t$ where all pairs $x_iy_i$
correspond to overlapping subsets.  We assume the standard communication 
model with two players, Alice and Bob, where Alice receives $x_1 \cdots x_t$
and Bob receives $y_1 \cdots y_t$.  (Observe that $(L_r)^t$ has small NFA's
with poly$(r+t)$ states.)
\begin{fact} \label{fact}
(\cite{HS}, pages 51-53).  Let $r,t \in \mathbb{N}$ be arbitrary.
If a deterministic protocol $D$ accepts only strings from $(L_{r})^t$
and if at most $2^{\alpha \cdot r \cdot t}$ messages are exchanged,
then $D$ accepts at most  $|(L_{r})^t|/2^{\alpha \cdot t} $ strings
from $(L_{r})^t$.  ($\alpha$ is a sufficiently small constant).
\end{fact}
Of particular interest are FNA's, for instance since their
equivalence problem is efficiently solvable.  However a separation
of FNA's and PNA's has remained open for almost twenty years
\cite{L98,RI}.  We are able to show such a separation and even prove
a hierarchy result for polynomial ambiguity.  To describe our result
we introduce the languages used in the separation.  For a language
$L$ of strings of identical length define
\begin{eqnarray*} \exists_{k}(L) & = & \{ w_1 w_2 \cdots  w_m  \; |
\; m \in \mathbb{N} \mbox{ and } w_{i} \in L \mbox{ for at least $k$
different positions} \}.
\end{eqnarray*} Thus the input is partitioned into blocks of identical
length and an input is accepted iff at least $k$ blocks belong to the
finite set $L$.  Now assume that $L$ can be recognized by a small NFA
$N$. Since $L$ is a finite set, we can recognize $\exists_k (L)$
by an NFA with ambiguity $O(n^k)$, if we increase the size of $N$
by at most the factor $k$.

How should the languages $L$ look like? In a first attempt set
$L = \{uv \;|\; u,v \in \{0,1\}^r, u \not= v \}$ as the language
of inequality between $r$-bit strings. Then $L$ is recognizable by
an NFA with ${\rm{poly}}(r)$ states and (bounded) ambiguity $r$. But
$\exists_1(L)$ is also recognizable with ${\rm{poly}}(r)$ states
and ambiguity $r$: guess a position $i \in \{1, \ldots, r\}$ and
accept $u^1 v^1 \cdots u^m v^m$ if $u^j_i \not= v^j_i$ for some $1
\leq j \leq r$.

What went wrong?  Few advice bits suffice and these advice bits can
be remembered.  In our second (and successful) attempt we therefore
set $L = (L_r)^t$, where we work with the iterated language of non-disjointness
from Fact \ref{fact}. This construction has two advantages. Firstly,
$L$ has a small NFA. Secondly, at least
intuitively, the number of guesses required for $L$ increases
exponentially with $t$ and hence a small NFA's for
$\exists_1(L_r)$ cannot remember sequences of $t$ guesses.  Our main result
verifies this intuition.

\begin{theorem} \label{main} Let $r \in \mathbb{N}$ be arbitrary. Set
$t=r^{1/3}$ and $L=(L_{r})^t$. Any NFA
for $\exists_k (L)$ with ambiguity $o(n^k)$ has at least
$2^{\Omega((r/k^2)^{1/3})}$ states. However, $\exists_k (L)$ can
be recognized by an NFA with ambiguity $O(n^k)$ and size $k \cdot
{\rm{poly}} (r)$.
\end{theorem}  Observe that we have obtained the claimed separation
of FNA's and PNA's for $k=1$, but Theorem \ref{main} also establishes
a hierarchy of polynomial ambiguity.

\section{A Proof Sketch} \label{sketch}

We start by proving Theorem \ref{main} for $k=1$.  Let $L = (L_r)^t$
and assume that the NFA $N$ recognizes $\exists_1 (L)$ with sublinear
ambiguity.  Observe that all strings in $L$ have length $2t$ and hence
strings in $\exists_1(L)$ have blocks of identical length $2t$.  We set
$K = \Sigma_r^{2t}$, where $\Sigma_r$ is the alphabet of $L_r$.  Finally set
\begin{eqnarray*} \exists_{=0} (L)& = & \{ w_1 w_2 \cdots
w_m : m \in \mathbb{N} \mbox{ and $w_i \in K \setminus L$ for all
$i$ } \}.
\end{eqnarray*}  Thus, as in the definition of $\exists_1 (L)$, the
input is partitioned into blocks and an input is accepted iff
\emph{no} block belongs to the finite set $L$. The computationally
hardest task for the NFA $N$ is to separate $\exists_{=0}(L)$ from
$\exists_1(L)$.

The critical part of the argument is to exploit the limitation of
sublinear ambiguity.  Let $Q$ be the set of states of $N$.
In Section \ref{autocom} we construct states $p_0, p_1 \in Q$ such
that at least $|L|/|Q|^2$ strings in $L$ have a computation
starting in $p_0$ and ending in $p_1$.  Moreover we show in Lemma
\ref{densit} that for any string $z' \in K \setminus L$ there is a
string $u \in \exists_{=0} (L)$ such that strings $S(z')$ with period
$z'u$ can be ``stored'' in a ``launching cycle'' before reaching
$p_0$ and in a ``storage cycle'' after leaving $p_1$.  The launching
cycle has the form $r \stackrel{(z' u)^a}{\to} r$ and allows to reach
$p_0$ via a computation $r \stackrel{(z' u)^{a_1}}{\to} p_0$;
analogously the  storage cycle is built from computations $ p_1
\stackrel{(u z')^{a_2}}{\to} s$ and $s \stackrel{(u z')^{a}}{\to} s$.
So far the launching cycle is harmless, since it delivers strings in
$\exists_{=0} (L)$ to state $p_0$, but these strings cannot use
computations from $p_0$ to $p_1$ which may be reserved for strings
in $L$. However, if a single occurrence of $z'$ within $S(z')$ is
replaced by an impostor string $z \in L$ and if the launching cycle
does not detect the replacement, then $N$ is forced into linear
ambiguity, provided the impostor $z$ can also hide at a matching
position within the storage cycle (see Lemma
\ref{1}).

Thus the NFA $N$ has to solve the ``detection problem'', namely it
has to detect whether an impostor $z \in L$ has replaced an
occurrence of $z' \in K \setminus L$ in both cycles.
The detection problem is set up in such
a way that \begin{itemize} \item[-] at least $|L|/|Q|^2$ strings
from $L$ are accepted, namely those strings $z \in L$ with a
computation $p_0 \stackrel{z}{\to} p_1$, and \item[-] all strings $z$
which for some $z' \in K \setminus L$ survive in matching positions within both cycles are
rejected.  In particular, all strings in $K \setminus L$ are rejected,
since a string $z \in K \setminus L$ is its own impostor.
\end{itemize}  Observe that no string $z$ is simultaneously accepted as
well as rejected, since all impostors have to be detected.  $N$ may
try to solve the detection problem unconventionally for instance by
allowing a potential impostor $z$ to survive undetected within the launching
and storage cycle, but not allowing $z$ to survive in
\emph{matching positions} within both cycles. Also $N$ does not have
to solve the detection problem completely, since
it can tolerate an impostor $z$ without a computation $p_0
\stackrel{z}{\to} p_1$.

We then simulate $N$ in Section \ref{dcom} by a nondeterministic
communication protocol which rejects all strings in $K \setminus L$,
accepts at least $|L|/|Q|^2$ strings in $L$ and does not
simultaneously accept and reject a string in $K \setminus L$ (see Lemma
\ref{com}).  Thus we have reduced the problem of avoiding linear
ambiguity for NFA's recognizing $\exists_1 (L)$ to a communication
problem in which a rather small minority of strings in $L$ has to be
separated from all of $K \setminus L$.   We show in Lemma \ref{det} how
to transform such a nondeterministic protocol into a deterministic protocol
by increasing the number of messages only subexponentially.  We are left
with a deterministic protocol which rejects all strings in $K \setminus L$
and accepts at least $|L|/|Q|^2$ strings in $L$. Finally the argument concludes
with an application of Fact \ref{fact}.  Thus, as in the case of exponential
ambiguity, we again have reduced the conciseness problem to an
investigation of deterministic protocols which recognize a ``small,
but significant chunk'' of a given product language.

The general case of ambiguity $O(n^k)$ is tackled in Section
\ref{hierarchy}. Showing the existence of launching and storage
cycles has now become a more complex problem. Previously it was
sufficient that the periodic string $S(z)$ was ``living'' in the one
launching and the one storage cycle. Now we have to work with a
vector $p_0,p_1, \ldots ,p_{2k-2},p_{2k-1}$ of states and have to
move $S(z)$ to $p_0$ and all the way from $p_{2i+1}$ to $p_{2(i+1)}$
for all $i=0, \ldots k-2$ and finally from $p_{2k-1}$ to an
accepting state.

\section{From Automata to Communication} \label{autocom}

We begin by utilizing the special structure of the languages
$\exists_1 (L)$.

\begin{definition}
Let $N$ be an NFA for $\exists_1(L)$ with initial state $q_0$. Let
$p$ be an arbitrary state of $N$.
\begin{itemize}
\item[(a)] We say that a string $v \in \exists_{=0}(L)$
\textbf{reaches} state $p$ iff there is a string $u \in
\exists_{=0}(L)$ and a computation for $u \cdot v$ which starts in
$q_0$ and ends in $p$. Moreover state $p$ \textbf{accepts} $v \in
\exists_{=0}(L)$ iff there is a string $w \in \exists_{=0}(L)$ and
an accepting computation for $v \cdot w$ starting in $p$.
\item[(b)] A pair $(p_0,p_1)$ of states of $N$ is \textbf{critical}
for the pair $(\xi_0, \xi_1) \in \exists_{=0}(L) \times
\exists_{=0}(L)$ iff all strings in $\exists_{=0}(L) \cdot \xi_0$
reach $p_0$ and all strings in $\xi_1 \cdot \exists_{=0}(L)$ are
accepted by $p_1$.
\end{itemize}
\end{definition}  Our next goal is to construct a pair $(\xi_0, \xi_1)
\in \exists_{=0}(L) \times \exists_{=0}(L)$ such that for all
strings $u \xi_0 z \xi_1 w \in \exists_{=0} (L) \cdot (\xi_0 \cdot L
\cdot \xi_1) \cdot \exists_{=0}(L)$ acceptance  is ``decided'' by
critical pairs. In particular we construct $(\xi_0, \xi_1)$ such
that there are accepting computations of the form $q_0 \stackrel{u
\xi_0 }{\to} p_0 \stackrel{z}{\to} p_1 \stackrel{\xi_1 w }{\to} q_f$
for a final state $q_f$ and a critical pair $(p_0,p_1)$ for
$(\xi_0,\xi_1)$. The crucial advantage of a critical pair is that
all strings in $\exists_0(L) \cdot \xi_0$ reach $p_0$ and all
strings in $\xi_1 \cdot \exists_0(L)$ are accepted by $p_1$; in
particular, there is no transition $p_0 \stackrel{z}{\to} p_1$ for a
string $z \in \exists_0(L)$ and acceptance is indeed decided by
$(p_0,p_1)$.

\begin{lemma} \label{onetwo} Let $N$ be an NFA for $\exists_1(L)$.
Then there are strings $\xi_0, \xi_1 \in \exists_{=0}(L)$ such that
$$ \bigcup_{(p_0,p_1) \;{\rm{is}}\; {\rm{critical}} \; {\rm{for}}\;
(\xi_0, \xi_1)} \{ z \in L \;|\; p_0 \stackrel{z}{\to} p_1 \} = L .$$
\end{lemma}

\proof  We process the states of $N$ in two phases.  In the first
phase we construct a string $\xi_0 \in \exists_{=0}(L)$ such that each
state $p$ is either \emph{alive} for $\xi_0$ (i.e., all strings in
$\exists_{=0}(L) \cdot \xi_0$ reach $p$) or \emph{dead} for $\xi_0$
(i.e., no string in $\exists_{=0}(L) \cdot \xi_0$ reaches $p$). The
construction process proceeds iteratively by processing all states $p$
of $N$ in an arbitrary order. We begin by setting $\xi_0 = \epsilon$.
When processing state $p$ we differentiate two cases.

{\bf Case 1:} All strings in $\exists_{=0}(L) \cdot \xi_0$ reach
$p$. We do not modify $\xi_0$.  Observe that  $p$ is alive for
$\xi_0$ and stays alive for any string in $\exists_{=0}(L)$ with
suffix $\xi_0$.

{\bf Case 2:}  There is a string $\xi \in \exists_{=0}(L)$ such that
$\xi \cdot \xi_0$ does not reach $p$.  The string $\xi \cdot \xi_0$
does not reach $p$ and hence no string in $\exists_{=0}(L) \cdot \xi
\cdot \xi_0$ has a computation beginning in the starting state $q_0$
and ending in $p$.  We replace $\xi_0$ by $\xi \cdot \xi_0$ and $p$
is dead for $\xi_0$, but also dead for any string in $\exists_{=0}
(L)$ with suffix $\xi_0$. Also observe that any already processed
state $q$ stays alive, resp. remains dead.

In the second phase we proceed completely analogously, but now
construct a string $\xi_1 \in \exists_{=0}(L)$ such that each state
$p$ is either \emph{alive} for $\xi_1$ (i.e., $p$ accepts all
strings in $\xi_1 \cdot \exists_{=0}(L) $) or \emph{dead} for $\xi_1$
(i.e., $p$ does not accept any string in $ \xi_1 \cdot \exists_{=0}(L)
$).

Now consider any string $s = \xi_0z\xi_1$ in $M = \xi_0 \cdot L
\cdot \xi_1$. Observe that $M$ is a subset of $\exists_1(L)$.
However $\xi_0$ cannot reach a dead state for $\xi_0$ and $\xi_1$
cannot be accepted by a dead state for $\xi_1$. Thus any accepting
computation for $s$ has to utilize a transition $p_0
\stackrel{z}{\to} p_1$ between alive states $p_0$ for $\xi_0$ and
$p_1$ for $\xi_1$. But any pair $(p_0,p_1)$ of alive states is a
critical pair and we are done. \qed

From now on we fix a pair $(\xi_0, \xi_1)
\in \exists_{=0}(L) \times \exists_{=0}(L)$ for
which Lemma \ref{onetwo} holds.  Let $(p_0,p_1)$ be an arbitrary
critical pair for $(\xi_0, \xi_1)$. We now utilize that all strings
in $\exists_{=0} (L) \cdot \xi_0$ reach $p_0$ and all strings in
$\xi_1 \cdot \exists_{=0}(L)$ are accepted by $p_1$.

\begin{lemma} \label{densit}  For all strings
$z \in K \setminus L$ there are states $r,s$, integers $a \geq 1,
a_1, a_2$ (with $a_1 + a_2 \leq a$) and a string $u \in
\exists_{=0}(L)$ as well as computations
\begin{eqnarray} \label{one}
& & r \stackrel{(z u)^a}{\to} r
\stackrel{(z u)^{a_1}}{\to} p_0 \; \mbox{ {\rm{and}}} \\
\label{two} & & p_1 \stackrel{(u z)^{a_2}}{\to} s \stackrel{(u
z)^{a}}{\to} s.
\end{eqnarray}
\end{lemma}
\proof We consider all strings of the form
$$\alpha(z) = (z \xi_1 \xi_0)^{|Q|} \; \mbox{ and } \;
\beta(z) = (\xi_1 \xi_0 z)^{|Q|}.$$ The string $\alpha(z)$ has suffix
$\xi_0$ and hence $\alpha(z)$ reaches $p_0$.  As a consequence there
is $\xi \in \exists_{=0}(L)$ and a computation $C$ for $\xi \cdot
\alpha(z)$ which begins in the initial state $q_0$ and reaches
$p_0$. After reading $\xi$, computation $C$ processes $\alpha(z)$
and produces a sequence of $|Q|+1$ states, where we list all states
before reading a copy of $z \xi_1 \xi_0$, resp. after reading the
last copy. A state $r$ of $N$ appears twice in this sequence and we
obtain a transition of the form $r \stackrel{(z \xi_1 \xi_0)^a}{\to}
r$ for $a \geq 1$. Finally $C$, starting in $r$, reaches $p_0$ after
reading the remaining $a_1$ copies.

To establish (\ref{one}), we set $u = \xi_1 \xi_0$ and obtain
transitions $r \stackrel{(z u)^a}{\to} r$ and $r
\stackrel{(zu)^{a_1}}{\to} p_0$. Thus (\ref{one}) follows. Part
(\ref{two}) is established by a similar argument, but now applied to
$\beta(z)$. This time we get  transitions $p_1 \stackrel{(uz
)^{a_2}}{\to} s$ and $s \stackrel{(uz)^{b}}{\to} s$.  But then $r
\stackrel{(z u)^{ma}}{\to} r$ as well as $s \stackrel{(u
z)^{m'b}}{\to} s$ are transitions for any multiples $m, m' \geq 1$
and the claim follows, if we replace both $a$ and $b$ by $ab(a_1 +
a_2) \geq a_1 + a_2$. \qed

Let $(p_0,p_1)$ be a critical pair for $(\xi_0,\xi_1)$.
We now introduce the detection problem for $(p_0,p_1)$ in which strings in $L$
have to be ``weakly'' separates  from strings in $K \setminus L$.
It turns out that any NFA $N$ for $\exists_1(L)$ solves the detection
problems for all critical pairs, provided $N$ has ambiguity $o(n)$.
Since we show later that $N$ can be efficiently simulated by a
communication protocol --with communication resources related to the
number of states-- and that the detection problem is hard for
communication complexity, $N$ must have many states.
The detection problem of $(p_0,p_1)$ has the following form:
\begin{itemize} \item[(a)]
Accept a string $z \in K$ iff there is a computation $p_0
\stackrel{z}{\to} p_1$ of $N$. Remember that for no $z \in K
\setminus L$ there is a computation
$$q_0 \stackrel{\xi \xi_0}{\to} p_0 \stackrel{z}{\to} p_1
\stackrel{\xi_1 \xi'}{\to} {q_f}$$ with the initial state $q_0$, a
final state $q_f$ and strings $\xi, \xi', \xi_0, \xi_1 \in
\exists_{=0}(L)$. Hence no string $z \in K \setminus L$ is accepted.

\item[(b)]  Reject a string $z \in K$ iff there are states
$r,r',r'',s,s',s''$, integers $a \geq 1, a_1, a_2$ (with $a_1+a_2
\leq a$) and strings $u \in \exists_{=0}(L), z' \in K \setminus L$
with computations
\begin{eqnarray} \label{xx1}
& & r \stackrel{(z'u)^{a_1}}{\to} r' \stackrel{zu}{\to} r''
\stackrel{(z' u)^{a - a_1-1}}{\to} r \stackrel{(z' u)^{a_1}}{\to}
p_0 \; \mbox{ and }\; \\ & & p_1 \stackrel{(u z')^{a_2}}{\to} s
\stackrel{(uz')^{a-a_2-1}}{\to} s' \stackrel{u z}{\to} s''
\stackrel{(uz')^{a_2}}{\to} s. \label{xx2}
\end{eqnarray} (The computations (\ref{xx1}) and (\ref{xx2}) will be used later to
define a launching and storage cycle respectively.  It turns out
that $z$ is placed within matching positions of the $z'u$- and $uz'$-cycle
and hence $z$ plays the role of an impostor of $z'$.)
\item[(c)] $z \in K$ is left undecided iff $z$ is neither accepted
nor rejected. \end{itemize} To explain the purpose of these transitions
consider the string
\begin{eqnarray*} S_1  = [zu \cdot (z'u)^{a-1}] \cdot [zu \cdot
(z'u)^{a-1}].
\end{eqnarray*} If we process the first half $zu \cdot (z'u)^{a-1}$
of $S_1$ starting in state $r'$, then there is a computation $C_0$
of the form $$r' \stackrel{zu}{\to} r'' \stackrel{(z' u)^{a -
a_1-1}}{\to} r \stackrel{(z'u)^{a_1}}{\to} r'$$ as well as a
computation $C_1$ from $r'$ to $p_0$ according to (\ref{xx1}). When
reading the second half of $S_1$, computation $C_0$ splits into a
computation $C_{00}$ which goes full circle reaching state $r'$
again and a computation $C_{01}$ which reaches $p_0$ after
completely reading $S_1$.  Now assume that there is a transition
$p_0 \stackrel{z}{\to} p_1$. Computation $C_1$ has reached $p_0$
after reading the first half of $S_1$ and now reads the second half
$zu \cdot (z'u)^{a-1} = z \cdot (uz')^{a-1} \cdot u$ of $S_1$. It
travels from $p_0$ to $p_1$ and subsequently reaches state $ s''$,
if \emph{additionally} the string $z$ is read. We have been
successful
\begin{itemize}
\item[(1)] in ``storing'' a mother computation via computation
$C_{00}$ in state $r'$,
\item[(2)] preparing for a new ``launch'' in state $p_0$ via
computation $C_{01}$ and \item[(3)] ''storing'' offspring
computations in state $s''$ via computation $C_1$.
\end{itemize}
We utilize properties (1)-(3) by defining a sequence $(S_m \;|\; m
\geq 1)$ with many computations, namely we set
\begin{eqnarray*} S_{m+1} =  S_m \cdot [ z u \cdot (z'u)^{a-1}]
= S_m \cdot [z \cdot (uz')^{a-1} \cdot u].
\end{eqnarray*} Assume inductively that there are computations for
$S_m$ which have reached the states $r'$ and $p_0$ respectively and
a computation for $S_m \cdot z$ which has reached $s''$. After
reading the suffix $zu \cdot (z'u)^{a-1}$ of $S_{m+1}$, the
computation starting in $r'$ has split into a computation reaching
$r'$ again and a computation reaching $p_0$, whereas the freshly
launched computation reaches $s''$ from $p_0$ after reading $S_{m+1}
\cdot z$.  Observe that all previously launched computations go full
circle after reading $(uz')^{a-1} \cdot uz$ and again have reached
state $s''$.  As a consequence, there are $m$ distinct computations
for $S_m z$ all reaching state $s''$ at the same time.

We say that $N$ has no redundant states, if each state is part of some
accepting computation of $N$. Which strings are rejected and which strings
are accepted?
\begin{lemma}  \label{1}  Let $N$ be an NFA recognizing
$\exists_1(L)$ without redundant states. Also assume that
$N$ has ambiguity $o(n)$.
\begin{itemize}
\item[(a)] Consider the detection problem of an arbitrary critical pair
$(p_0,p_1)$. Then all strings in $K \setminus L$ are rejected and no string in
$K$ is simultaneously accepted and rejected.
\item[(b)] Each string in $L$ is accepted in the detection problem of
some critical pair.
\end{itemize}
\end{lemma}

\proof (a) We observe first that every string $z \in K \setminus L$ is rejected.
Why? We may choose $z'=z$ and the transitions required in (\ref{xx1}) and (\ref{xx2})
exist as a consequence of Lemma \ref{densit}: the states $r',r''$ and $s',s''$
belong to the $r$-cycle and the $s$-cycle respectively.

Now assume that there is a string $z \in K$ which is
accepted and rejected.  Since $z$ is accepted, there is a
computation $p_0 \stackrel{z}{\to} p_1$. Also, since $z$ is
rejected, there are computations of the form (\ref{xx1}) and
(\ref{xx2}). Thus we may construct the strings $S_mz$ for every $m$
and obtain $m$ distinct computations which, starting from state $r'$,
reach state $s''$ at the same time. But $N$ does not have redundant
states and  each state, and in particular state $r'$, is reachable
from the initial state.  Also each state, and in particular state
$s''$, can reach an accepting state. Thus there are strings $\xi_0, \xi_1$
such that $\xi_0 \cdot S_m z \cdot \xi_1$ has $m$ accepting computations.
But $S_mz$ is a string with length linear in $m$ and hence $N$ has at least
linear ambiguity.

(b) follows from part (a), if we apply Lemma \ref{onetwo}. \qed

\section{The Communication Problem} \label{dcom}

We show that the detection problem has an efficient
communication protocol, provided a small NFA $N$ with ambiguity $o(n)$
recognizes $\exists_1(L)$. Remember that $L= (L_r)^t$ and $K = \Sigma_r^{2t}$.
We work with the conventional two-party communication model consisting of two
players Alice and Bob. If $x_1 y_1 \cdots x_t y_t$ is the input of $N$,
then Alice receives $x_1 \cdots x_t$ and Bob receives $y_1 \cdots y_t$ as
their respective inputs. Alice and Bob communicate nondeterministically with
computations either being accepting, rejecting or undecided. We say
that an input is accepted if at least one computation is accepting,
rejecting if at least one computation is rejecting and undecided if
all computations are undecided.  (Thus undecided computations play
the role of rejecting computations for conventional nondeterminism.)
Observe that we allow to simultaneously accept and reject an input.

Now assume that the NFA $N$ recognizes $\exists_1(L)$.  Let $q,q^*$
be two states of $N$ and let $z \in K$ be an input string.  Our
first goal is to determine whether $N$ has a computation for $z$
starting in $q$ and ending in $q^*$. Set $q_0 = q$.  Beginning with
$i=1$, Alice simulates $N$ for input $x_{i}$ by starting in
state $q_{i-1}$ and sends state $q_i'$, if $q_{i}'$ is reached.
Bob simulates $N$ for input $y_{i}$ by starting in state
$q_{i}'$ and sends state $q_{i}$, if $q_i$ is reached.  In the last
round Bob accepts if additionally $q_{s}=q^*$ holds and otherwise
outputs ``undecided''. Obviously the simulating protocol
exchanges at most $|Q|^{2t}$ messages.  It has an accepting
computation iff $N$ has a computation $q \stackrel{z}{\to} q^*$ and
otherwise leaves the input undecided.

We say that a protocol solves the detection problem of $(p_0,p_1)$
if the protocol labels each input as accepted, rejected or undecided
as prescribed by the detection problem.

\begin{lemma}  \label{com}
Assume that $N$ recognizes $\exists_1(L)$ and that $N$ has ambiguity
$o(n)$. Let $(p_0,p_1)$ be a critical pair for $(\xi_0,\xi_1)$. Then
there is a nondeterministic protocol $P$ which solves the
detection problem of $(p_0,p_1)$ with $|Q|^{O(t)}$ messages.
\end{lemma}

\proof  We begin by describing the protocol $P$.
In its first attempt $P$ tries to accept its input $z \in K$ by simulating
the automaton $N$ when reading $z$ starting in state $p_0$.
$P$ accepts $z$ iff state $p_1$ is reached and otherwise
leaves $z$ undecided.

In its second attempt $P$ tries to reject $z$. Alice guesses states
$r,r',r'',s,s',s''$ as well as strings $z' \in K \setminus L, u \in
\exists_{=0}(L)$ and integers $a_1,a_2,a$ (with $a_1 + a_2 \leq a$).
Then Alice verifies the following transitions \emph{without}
communication, namely
\begin{itemize}
\item[-] $r \stackrel{(z'u)^{a_1}}{\to} r'$ as well as
$r'' \stackrel{(z' u)^{a - a_1-1}}{\to} r \stackrel{(z'
u)^{a_1}}{\to} p_0$ and
\item[-] $p_1 \stackrel{(u z')^{a_2}}{\to} s
\stackrel{(uz')^{a-a_2-1}}{\to} s'$ as well as $s''
\stackrel{(uz')^{a_2}}{\to} s$.
\end{itemize} In order to check the remaining transition
$r' \stackrel{zu}{\to} r''$ and $s' \stackrel{u z}{\to} s''$,
Alice guesses additional states $\rho,\sigma$ and verifies
the transitions $\rho \stackrel{u}{\to} r''$ and $s'
\stackrel{u}{\to} \sigma$ by herself. Subsequently Alice
communicates the states $r', \rho$ as well as $\sigma,s''$ and both
Alice and Bob simulate the automaton $N$ on input $z$ for starting
states $r'$ and $\sigma$. Bob rejects iff the transitions
$r' \stackrel{z}{\to} \rho$ and $\sigma \stackrel{z}{\to} s''$ have
been verified and otherwise labels $z$ as undecided. Observe that
$P$ exchanges at most $|Q|^{O(t)}$ messages, since $P$ uses messages
only when simulating $N$ on the string $z \in K$. \qed

\section{From Nondeterminism to Determinism} \label{ncom}

In Lemma \ref{com} we have solved the detection problem of a
critical pair by a nondeterministic protocol $P$ with only $|Q|^{O(t)}$
messages. However the detection problem separates $L$ from its complement
$K \setminus L$ only weakly, since the majority of strings from $L$ are
either rejected or left undecided. We begin our analysis by transforming
the nondeterministic protocol $P$ into a deterministic protocol $D$.
We avoid an exponential blowup in the number of messages by observing
the structural limitations of $P$.  In particular, $P$ accepts a subset
$L_{\rm{yes}} $ of $ L$ and rejects a superset $L_{\rm{no}}$ of
$K \setminus L$, where $L_{\rm{yes}}$ and $L_{\rm{no}}$ are disjoint.

\begin{lemma} \label{det} There is a deterministic protocol $D$
which accepts at least $|L|/|Q|^2$ strings from $L$ and rejects all
strings from $K \setminus L$. No string is left undecided and no string
is accepted as well as rejected.  Moreover, at most $|Q|^{O(t^2
\cdot \log_2 |Q|)} $ messages are exchanged.
\end{lemma}

\proof We begin by fixing a critical pair $(p_0,p_1)$
such that at least $|L|/|Q|^2$ strings are accepted in the detection
problem of $(p_0,p_1)$.  Observe that such a critical pair exists
as a consequence of Lemma \ref{1} (b), since each string in $L$ is
accepted in the detection problem of at least one critical pair and
there are at most $|Q|^2$ critical pairs.

Let $L_{\rm{yes}} $ be the subset of $L$ which is accepted in the detection
problem of $(p_0,p_1)$ and let $L_{\rm{no}}$ be the superset of $K \setminus L$
of rejected strings. According to Lemma \ref{com} there is a nondeterministic
protocol $P$ which solves the detection problem of $(p_0,p_1)$ with at most
$|Q|^{O(t)}$ messages.  Thus there are conventional nondeterministic protocols
$P_{\rm{yes}}$ for $L_{\rm{yes}}$ and $P_{\rm{no}}$
for $L_{\rm{no}}$ which exchange at most $|Q|^{O(t)}$ messages each.

To obtain a deterministic protocol $D$ from
$P_{\rm{yes}}$ and $P_{\rm{no}}$ we utilize that
deterministic protocols with $M^{O(\log_2 M)}$ messages can be built
from nondeterministic protocols, provided the protocols recognize a
language \emph{and} its complement by exchanging at most $M$
messages \cite{aho}. Our situation however is more complicated,
since $L_{\rm{yes}}$ is only a subset of the complement of
$L_{\rm{}no}$. We employ the construction in \cite{Lo} with the
following modifications. Define the communication matrix $C$ of
$(P_{\rm{yes}},P_{\rm{no}})$ by setting
$$C [x_1 \cdots x_t, y_1 \cdots y_t] = \left\{ \begin{array} {ll}
1 & \;\; x_1y_1 \cdots x_ty_t \in L_{\rm{yes}}, \\ 0 & \;\; x_1y_1
\cdots x_ty_t \in L_{\rm{no}} \\ {\rm{undecided}} &
\;\;{\rm{otherwise.}}
\end{array} \right. $$
Each message $m$ corresponds to a submatrix $M$ of $C$ defined by the
collection of rows for which the message is sent and the collection of
columns for which it is accepted.  Now let $M$ be a submatrix of the
communication matrix $C$. We define $\Delta_{\rm{yes}}(M)$ to be the
maximal size of a submatrix $T$ of $M$, where $T$, after a suitable
permutation of rows and columns of $M$, is a lower triangular matrix
with ones on the diagonal and zeroes above the diagonal. (Observe that
$T$ may contain undecided entries, but these entries have to appear
below the diagonal.) Since $L_{\rm{yes}}$ is accepted by the nondeterministic
protocol $P_{\rm{yes}}$ and since no two diagonal entries can be
accepted by the same message, we obtain that $\Delta_{\rm{yes}}(C)$
is bounded by the number of messages of $P_{\rm{yes}}$ and
hence $\Delta_{\rm{yes}}(C) \leq |Q|^{O(t)}$ follows.

We first try to reject the given input by deterministically
selecting a sequence $m_i$ of messages from the protocol
$P_{\rm{no}}$.  As for the conventional transformation to
deterministic protocols, the triangular message complexity will
be halved in each step and in particular
$\Delta_{\rm{yes}}(M_1 \cap \cdots \cap M_i) \leq
\Delta_{\rm{yes}}(C)/2^i$ follows.  We proceed as in the conventional
transformation and stop the
communication prematurely,  if the output ``no'' can be
excluded and output ``yes''. Otherwise, after at most $\log_2
\Delta_{\rm{yes}}(C)$ rounds, we obtain
$\Delta_{\rm{yes}}(M_1 \cap \cdots \cap M_i) \leq 1$. As a consequence,
the submatrix $M_1 \cap \cdots \cap M_i$ has no triangular submatrix of
size two or larger.  In particular, the submatrix $M$ of $M_1 \cdots M_i$
spanned by all rows and columns of $M_i^*$ with a one, contains all
ones of $M_1 \cdots M_i$, no zeroes and possibly undecided entries.
If the joint input belongs to $M$, then we stop and accept, resp.
stop and reject otherwise. In each round only messages of
$P_{\rm{no}}$ and hence at most $|Q|^{O(t)}$ messages are
exchanged.  Thus overall at most $\big[ |Q|^{O(t)} \big]^{ \log_2
\Delta_{\rm{yes}}(C)} = |Q|^{O(t^2 \cdot \log_2|Q|)}$ messages are
generated. \qed

Remember that $L =(L_r)^t$, where $L_r$ is the language of non-disjointness
for $r$-element subsets of $\{1, \ldots , r^{32} \}$.  Let $D$ be a
deterministic protocol which accepts only strings in $L$.  Also let
$\alpha$ be a sufficiently small positive constant.  We apply Fact
\ref{fact} and obtain that $D$ accepts at most $|L|/2^{\alpha \cdot t} $
strings from $L$, provided at most $2^{\alpha \cdot r \cdot t}$ messages
are exchanged.

Now, if an NFA $N$ with sublinear ambiguity recognizes $\exists_1
(L)$, then we apply Lemma \ref{det} to obtain a
deterministic protocol which exchanges at most $|Q|^{O(t^2 \cdot
\log_2 |Q|)} $ messages, accepts at least $|L_r|/|Q|^2$ strings and
accepts only strings from $L$.  Thus, if $|Q|^{O(t^2
\cdot \log_2 |Q|)} \leq 2^{\alpha \cdot r \cdot t}$ for a
sufficiently small positive constant $\alpha$, then at most $|L|
/2^{\alpha \cdot t}$ inputs from $L$ are accepted.  But the
nondeterministic protocol accepts at least $|L|/|Q|^2$ strings
from $L$ and hence \begin{eqnarray}
\label{q} |Q| = 2^{\Omega(t)}
\end{eqnarray} follows.  We set $t = r^{1/3}$.  Let $\beta$ be a sufficiently
small positive constant. Now either $|Q| \geq 2^{\beta \cdot \sqrt{r/t}}$
and we are done, since then $|Q| = 2^{\Omega(r^{1/3})}$ or $|Q| < 2^{\beta
\cdot \sqrt{r/t}}$ holds. In the latter case
$$|Q|^{t^2 \cdot \log_2 |Q|} < 2^{(\beta \cdot \sqrt{r/t}) \cdot t^2
\cdot (\beta \cdot \sqrt{r/t})} = 2^{\beta^2 \cdot t \cdot r}$$ and
the upper bound on the number of messages in Fact \ref{fact} is met,
provided $\beta$ is sufficiently small.  But then $|Q| = 2^{\Omega(t)}$
follows from  (\ref{q}) and hence $|Q| \geq 2^{\gamma \cdot t}$ holds
for some positive constant $\gamma$.  We obtain $2^{\gamma \cdot t}
\leq |Q| < 2^{\beta \cdot \sqrt{r/t}} $ and hence $2^{\gamma \cdot t}
< 2^{\beta \cdot \sqrt{r/t}} = 2^{\beta \cdot t} $, since $t = r^{1/3}$.
We get a contradiction if $\beta$ is chosen sufficiently small and we have
shown
\begin{lemma} \label{final} Let $N$ be an NFA with sublinear
ambiguity recognizing $\exists_1 (L)$.  Then $N$ has at least
$2^{\Omega(r^{1/3})}$ states. \qed
\end{lemma}

\section{A Hierarchy for Polynomial Ambiguity} \label{hierarchy}

Let $k \geq 1$ be arbitrary and let $N$ be an NFA for $\exists_k(L)$.
We again follow the strategy for $k=1$,  however the transition from
NFA's to communication protocols is now more involved.
For $k >1$ we have to work with vectors $(p_0, p_1,
\ldots , p_{2(k-1)}, p_{2(k-1)+1})$ of states and besides
reachabilty for $p_0$ and acceptance by $p_{2(k-1)+1}$ we also have
to guarantee that computation paths exist between $p_{2i}$ and
$p_{2i+1}$. This last requirement requires some further work.

\begin{definition}
Let $(\xi_0,\xi_1) \in \exists_{=0}(L) \times \exists_{=0}(L)$ be
arbitrary.  We say that the vector $(p_0, p_1, \ldots , p_{2(k-1)},
p_{2(k-1)+1})$ is critical for $(\xi_0,\xi_1)$ iff
\begin{itemize}
\item[(1)] all strings in $\exists_{=0} (L) \cdot \xi_0$ reach $p_0$
and all strings in $\xi_1 \cdot \exists_{=0}(L)$ are accepted by
$p_{2(k-1)+1}$
\item[(2)] and for all strings $u \in \exists_{=0}(L)$  and for all $i$
($0 \leq i < k-1$) there is a string $v$ such that a computation for
$\xi_1 uv$ starts in $p_{2i+1}$ and ends in $p_{2(i+1)}$.
\end{itemize}
\end{definition}
We construct $\xi_0$ as in Lemma \ref{onetwo} and hence for any state
$p$ of the NFA $N$ either all strings in $\exists_{=0} (L) \cdot \xi_0$
reach $p$ or no such string reaches $p$.  To construct $\xi_1$ we first
run the procedure of Lemma \ref{onetwo} and property (1) is satisfied.
Then we process all pairs $(p,q)$ of states of $N$ in some arbitrary order.
If for all strings $u \in \exists_{=0} (L)$ there is a string $v \in
\exists_{=0} (L)$ such that $\xi_1 u v$ has a computation beginning
in $p$ and ending in $q$, then we say that the pair $(p,q)$ is
``alive'' and $\xi_1$ is left unchanged. Otherwise there is a string
$u \in \exists_{=0} (L)$ such that no computation for a string in
$\xi_1 \cdot u \cdot \exists_{=0} (L)$ has a computation beginning
in $p$ and ending in $q$.  We replace $\xi_1$ by $\xi_1 u$. The pair
$(p,q)$ is now ``dead'', since no string in $\xi_1 \cdot
\exists_{=0} (L)$ has a computation beginning in $p$ and ending in
$q$. Also observe that processed pairs do not change their status,
i.e., remain dead, resp. stay alive after updating $\xi_1$.
We have generalized Lemma \ref{onetwo}.

\begin{lemma} \label{onek} Let $N$ be an NFA for $\exists_k(L)$.
Then there are strings $\xi_0, \xi_1 \in \exists_{=0}(L)$ such that
$$ \bigcup_{(p_0, \ldots, p_{2k-1}) \;{\rm{is}}\; {\rm{critical}} \; {\rm{for}}\;
(\xi_0, \xi_1)} \{ z \in L \;|\; p_{2i+1} \stackrel{z}{\to}
p_{2(i+1)} \mbox{ for all  }0 \leq i < k-1 \} = L .$$
\end{lemma}  \proof The argument is analogous to the proof of Lemma
\ref{onetwo}.  This time we have to observe that accepting
computations for strings in $\xi_0 \cdot(L \cdot \xi_1)^{k}$  have
to traverse critical vectors. \qed

For $k=1$ Lemma \ref{densit} establishes that a string $S(z)$ ``lives''
in a launching cycle for $p_0$ and a storage cycle for $p_1$.  Its
generalization requires more work.  Let $\vec{p}= (p_0, \ldots, p_{2k-1})$
be a critical vector and let $z \in K \setminus L$ be an arbitrary
string. We construct a string $u \in \exists_{=0}(L)$ for $z$ so that
some string with period $uz$ can be launched by $p_0$, stored and launched
in between $p_{2i+1} $ and $p_{2(i+1)}$ and finally stored by $p_{2(k-1)+1}$.
In particular, we say that a string $u \in \exists_{=0}(L)$ is appropriate
for $z$ if the following properties are satisfied:
\begin{itemize}
\item[(1)] $(zu)^{|Q|}$ reaches $p_0$.
\item[(2)] For every $i$, $0 \leq i < k-1$, there is a string $s_i$ and
computations $p_{2i+1} \stackrel{s_i}{\to} p_{2(i+1)}$.  Moreover, $s_i$
starts with a suffix of $uz$ containing $\xi_1$ as prefix, followed by
$(uz)^{|Q|}$ and completed by a prefix of $u$.
\item[(3)] State $p_{2(k-1)+1}$ accepts any string $s_{k-1}$ which consists
of a suffix of $uz$ containing $\xi_1$ as prefix, followed by $(uz)^{|Q|}$.
\item[(4)] The string $s = (zu)^{|Q|} z s_0 z \cdots z s_{k-2} z s_{k-1}$
has periods $zu$ and $uz$ respectively.
\end{itemize}
Now assume that $u$ is appropriate for $z$.  We show that the string $S(z)=s$
``lives'' in appropriate cycles for each $p_i$. First observe that $S(z)$ has
period $zu$ and hence also period $uz$.  The proof of
Lemma \ref{densit} shows that a launching cycle $r \stackrel{(z u)^a}{\to} r
\stackrel{(z u)^{a_1}}{\to} p_0$ is established, once $(zu)^{|Q|}$ reaches
$p_0$.  Also, intermediate cycles in between $p_{2i+1} $ and $p_{2(i+1)}$
exist, since $s_i$ has substring $(uz)^{|Q|}$, and a final storage
cycle following $p_{2(k-1)+1}$ exists, since $p_{2(k-1)+1}$ accepts a string
with suffix $(uz)^{|Q|}$.
\begin{lemma} \label{twok}
For every string $z \in K \setminus L$ there is an appropriate string
$u \in \exists_{=0}(L)$ for $z$.
\end{lemma}
\proof Let $q_l$ be some arbitrary ordering of the states of $N$.  Each
pair $(p_{2i+1}, p_{2(i+1)})$ influences the construction of $u$.  Assume
for the moment that strings $\xi_{i,l}$ are already defined.  We set
$$ u_{i,j} = \xi_1 \cdot \Pi_{l \leq j, (q_l,p_{2(i+1)}) \;
{\rm{is}} \; {\rm{alive}}} \;\; z \xi_1 \xi_{i,l}$$ for all $j$ ($1
\leq j \leq |Q|$).  Observe that $u_{i,j} = u_{i,j-1} \cdot (z \xi_1
\xi_{i,j})$, if $(q_j,p_{2(i+1)})$ is alive, and that $\xi_1$ is a
prefix of $u_{i,j}$. Choose the strings $\xi_{i,l} \in \exists_{=0}(L)$
so that there is a computation for $u_{i,j}$ from $q_j$ to $p_{2(i+1)}$.
Such strings $\xi_{i,l}$ exist with property (2) of a critical vector,
since $\xi_1$ is a prefix of $u_{i,j}$ and $(q_j, p_{2(i+1)})$ is alive.
Finally set $$u_i = u_{i,|Q|} \cdot z \xi_1 \;\;\; {\rm{and}} \;\;\; u =
u_0 \cdots u_{k-2} \cdot \xi_0.$$ We show that $u$ is appropriate for $z$
by first verifying property (1). The string $u$ has suffix $\xi_0$ and
hence, by property (1) of a critical vector, $(zu)^{|Q|}$ reaches $p_0$,
the first component of the critical vector $\vec{p}$.

Observe that each $u_{i,j}$ has prefix $\xi_1$ and hence $u_i$ and $u$ have
$\xi_1$ as prefix.  We start the verification of properties (2) and (3) by
defining $s_0$ and constructing a computation $p_{1} \stackrel{s_0}{\to} p_{2}$.
Since $\xi_1$ is a prefix of $u$, there is a computation for $(uz)^{|Q|}$
which leads from $p_1$ to a state $q_j$ such that the pair $(q_j,p_2)$ is
alive. But then, by definition of $u_{0,j}$, there is a computation for
$(uz)^{|Q|} \cdot u_{0,j}$ which starts in $p_1$, reaches $q_j$ after reading
$(uz)^{|Q|}$ and ends in $p_2$ after reading $u_{0,j}$.  We set $s_0
= (uz)^{|Q|}\cdot u_{0,j}$.  By construction, $u_{0,j}$ is a prefix of
$u_0$ which itself is a prefix of $u$.  Thus there is a string $v_{0,j}$
with $u = u_{0,j} \cdot z \cdot v_{0,j}$ and $v_{0,j}$ has prefix $\xi_1$.

We now construct a string $s_1$ and a computation $p_3 \stackrel{s_1}{\to}
p_4$ as follows.  Since $v_{0,j}$ has prefix $\xi_1$ there is a computation for
$v_{0,j} \cdot z \cdot (uz)^{|Q|} \cdot u_0$ which reaches a state $q_k$
when starting in state $p_3$. Since the pair $(q_k,p_4)$ is alive,
we obtain the computation
\begin{eqnarray*} p_3 \stackrel{v_{0,j}z
(uz)^{|Q|}u_0}{\to} q_k \stackrel{u_{1,k}}{\to} p_4 \end{eqnarray*}
and set $s_1 = v_{0,j}z (uz)^{|Q|}u_0 u_{1,k}$.  The construction of
$s_i$ and verifying a computation $p_{2i+1} \stackrel{s_i}{\to} p_{2(i+1)}$
for arbitrary $i < k-1$ proceeds in a completely analogous fashion.
Finally, again by property (1) of a critical vector, state $p_{2(k-1)+1}$
accepts any string $s_{k-1}$ consisting of a suffix of $uz$ followed by
$(zu)^{|Q|}$, since the suffix of $uz$ has prefix $\xi_1$.

To complete the argument observe that by construction
$s = (zu)^{|Q|} z s_0 z \cdots z s_{k-2} z s_{k-1}$ has periods $uz$ and
$zu$ respectively.  \qed

The remainder of the argument proceeds completely analogous to the case of
$k=1$.  Lemma \ref{1} shows that an NFA with sublinear ambiguity solves the
detection problem for $k=1$. To introduce its generalization we firstly introduce
the detection problem for $k>1$: $z$ has to be
rejected iff there is a string $z' \in K \setminus L$  such that $z$, acting
as an impostor of $z'$, can be placed in matching positions within the $k+1$
individual $uz'$-cycles of $N$.  Lemma \ref{1} was a direct consequence of
Lemma \ref{densit} in the case of $k=1$.  In the same manner we can now show
that an NFA with ambiguity $o(n^k)$ solves the detection problem for general
$k$ as a direct consequence of Lemma \ref{twok}.

Let $N$ be an NFA with ambiguity $o(n^k)$ for $\exists_k(L)$.  As in Lemma
\ref{com} we simulate $N$ to obtain a nondeterministic protocol $P$ solving the
detection problem with $|Q|^{O(kt)}$ messages; the exponent grows by the factor
$k$, since $k+1$ instead of two computations of $N$ on input $z$ have to be
simulated.  We transform $P$ into a deterministic protocol $D$ with
$|Q|^{O((k t)^2 \log |Q|)}$ messages as in Lemma \ref{det}. To complete
the proof of Theorem \ref{main}, we replace $r$ by $r/k^2$ in the proof of Lemma
\ref{final} (to compensate for the increase in the number of messages of $D$
from $|Q|^{O(t^2 \log |Q|)}$ to $|Q|^{O(k^2 t^2 \log |Q|)}$) and obtain

\begin{lemma} \label{final2} Let $N$ be an NFA with ambiguity $o(n^k) $
recognizing $\exists_k (L)$.  Then $N$ has at least
$2^{\Omega((r/k^2)^{1/3})}$ states. \qed
\end{lemma}

\end{document}